\documentclass[11pt]{article}

\usepackage[T2A]{fontenc}
\usepackage[cp1251]{inputenc}
\usepackage[english,russian]{babel}
\usepackage{amsmath}
\usepackage{amsthm}
\usepackage{amssymb}
\usepackage{amscd}
\usepackage[dvips]{graphicx}
\usepackage{wrapfig}
\usepackage{titlesec}
\usepackage{color}
\usepackage{graphicx}
\usepackage{multicol}
\usepackage{floatflt}
\usepackage{indentfirst}

\theoremstyle{plain}
\newtheorem{Theorem}{Теорема}
\newtheorem{Lemma}{Лемма}

\theoremstyle{definition}
\newtheorem{Example}{Пример}
\newtheorem{Remark}{Замечание}

\newenvironment{Proof}{\par\noindent\textbf{Доказательство.}}{\hfill$\scriptstyle\blacksquare$\vspace{3mm}\par}

\DeclareMathOperator{\const}{const}

\renewcommand\Re{\mathop{\mathrm{Re}}}
\renewcommand\Im{\mathop{\mathrm{Im}}}
\renewcommand\le{\leqslant}
\renewcommand\ge{\geqslant}
\renewcommand\phi{\varphi}

\newcommand\gterm[1]{\mathcal{#1}}
\newcommand\muld[1]{\mathbf{#1}}
\newcommand\abs[1]{\left|#1\right|}

\newcommand\Abel[1]{\vec A({#1})}
\newcommand\ScalarP[2]{\left\langle #1,\, #2 \right\rangle}

\title{
    Функция Грина дискретного конечнозонного при одной энергии двумерного оператора Шредингера на квад-графе
    \thanks{Работа выполнена при поддержке гранта Правительства Российской Федерации 2010-220-01-077}
}
\author{
    Б.\,О.\,Василевский \thanks{МГУ им. М.\,В.~Ломоносова, email: vasilevskiy.boris@gmail.com}
}
\date{4 ноября 2013}

\begin{document}

\maketitle

\begin{abstract}
    Применяется конечнозонный подход для построения дискретного оператор Шредингера на квад-графе, представленном в виде двумерной целочисленной подрешетки в $d$-мерном пространстве. Функция Грина для этого оператора явно выражается в виде интеграла по специальных контурам от дифференциала, построенного по спектральным данным. Полученная функция имеет известную асимптотику.
\end{abstract}

\subsection{Введение}
    В теории интегрируемых систем математической физики важную роль играет конечнозонный подход. В непрерывном случае хотелось бы упомянуть работу Дубровина, Кричевера и Новикова~\cite{periodicShredinger}, в которой они показали интегрируемость двумерного стационарного конечнозонного оператора Шрёдингера при фиксированной энергии. Не только чисто теоретический интерес вызывает задача построения интегрируемых дискретизаций этого оператора. Интегрируемая гиперболическая дискретизация на квадратной решетке (построено обратное спектральное преобразование в периодическом случае) была найдена И.\,М.\,Кричевером~\cite{giperbolDiscr}. Далее, в статье А.\,Доливы, П.\,Гриневича, М.\,Нишпровски и П.\,Сантини~\cite{4authors} была получена эллиптическая дискретизация из специальной редукции гиперболической дискретизации. Эта редукция в терминах спектральных данных оказалась очень похожа на редукцию в работах Веселова и Новикова~\cite{finitShredinger} для непрерывного случая. В частности, на спектральной кривой требуется наличие голоморфной инволюции с двумя неподвижными точками.

    Случаи двух и нуля неподвижных точек у голоморфной инволюции на римановой поверхности являются наиболее интересными, согласно Д.\,Фэю~\cite{Fay}.
    Из работы Кричевера и Грушевского~\cite{KG} в частности следует, что общие решения отвечают инволюции без неподвижных точек, а решения из~\cite{4authors} являются специальными. Решения общего положения отвечают спектральным кривым, у которых инволюция не имеет неподвижных точек. Но вслед за~\cite{4authors} мы будем рассматривать инволюцию именно c двумя неподвижными точками.

    Хорошим обобщением квадратной решетки является квад-граф, у которого каждая грань по построению является четырехугольником. Здесь хотелось бы отметить работу А.\,Бобенко, К.\,Меркат, Ю\,Сурис~\cite{discran}. В одной из ее глав обсуждается отображение квад-графа в комплексную плоскость, при котором каждая грань переходит в параллелограмм, и интегрируемость возникающего при таком отображении оператора Коши-Римана в смысле <<3D-совместности>>. При этом случай положительных весов в точности соответствует уже ромбовидному вложению и является самым интересным.

    В настоящей статье делается шаг к применению конечнозонного подхода для построения дискретного оператора Шредингера $L$ на квад-графе. Функция Грина для $L$ явно выражается в виде интеграла по специальных контурам от дифференциала, построенного по спектральным данным. Полученная функция имеет известную асимптотику.

\subsection{Дискретные комплексы, квад-графы и комплексный анализ}
    Рассмотрим двумерный дискретный подкомплекс $\Omega_\gterm{D}$ $d$-мерной квадратной решетки $\mathbb{Z}^d$ для произвольного $d \ge 2$. Каждая грань $\Omega_\gterm{D}$ является двумерным единичным квадратом. Потребуем, чтобы подкомплекс укладывался $\mathbb{C}$ без самопересечений, то есть представлялся в виде планарного графа $\gterm{D}$. Обратное отображение отправляет вершины графа в узлы данной решетки $\muld{n}: V(\gterm{D}) \to \mathbb{Z}^d$. Каждая грань $\gterm{D}$ является четырехугольником, сам граф является двудольным.

    Возьмем вершины одной доли и соединим ребрами те из них, которые лежат в одной грани. Полученный граф обозначим через $\gterm{G}$, а построенный аналогично по другой доле --- $\gterm{G}^*$. Несложно видеть, что $V(\gterm{D}) = V(\gterm{G}) \sqcup V(\gterm{G}^*)$. Кроме того, для любой грани $\gterm{D}$ одна из ее диагоналей $e$ является ребром в $\gterm{D}$, а другая $e^*$ --- ребром в $\gterm{D}^*$. Наконец, $\gterm{G}$ и $\gterm{G}^*$ являются двойственными. Правильно выбрав поверхность $\Omega_\gterm{D}$, можно получить произвольный (планарный) граф $\gterm{G}$.

    Напомним некоторые определения из линейной теории дискретного комплексного анализа. Более подробное изложение можно найти например в~\cite{discran}. Пусть на ребрах графа $\gterm{G}$ определена комплекснозначная функция $\nu: E(\gterm{G}) \to \mathbb{C}$. Рассмотрим оператор $\Delta$, действующий на функциях $f: V(\gterm{G}) \to \mathbb{C}$ по формуле
    \begin{gather}\label{eqGenericLanlaceDef}
        (\Delta f)(x_0) = \sum\limits_{x \sim x_0} \nu(x_0, x)(f(x) - f(x_0)),
    \end{gather}
    где суммирование проходит по всем соседним с $x_0$ вершинам в графе $\gterm{G}$. Назовем этот оператор Лапласианом, соответствующим весовой функции $\nu$. Дискретной гармонической (относительно весовой функции $\nu$) назовем функцию функцию $f: V(\gterm{G}) \to \mathbb{C}$, для которой выполняется $\Delta f = 0$.

    Продолжим весовую функцию на ребра $e^* \in E(\gterm{G}^*)$ по формуле $\nu(e^*) = 1 / \nu(e)$, где $e \in E(\gterm{G})$ --- вторая диагональ соответствующего четырехугольника.
    \begin{figure}[ht]
        \centering
        \includegraphics{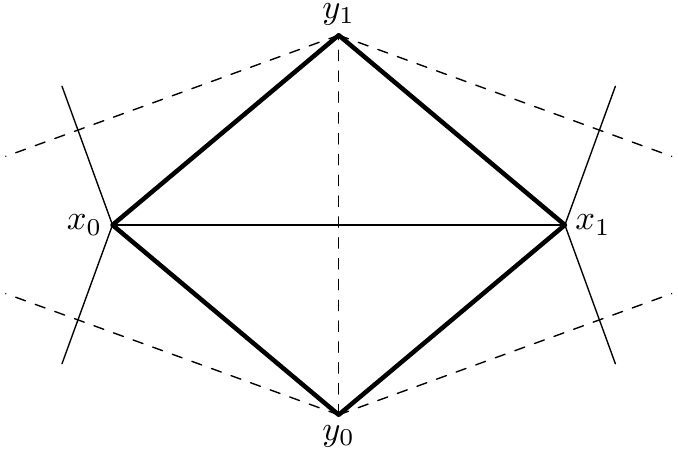}\\
        \caption{Грань $(x_0, y_0, x_1, y_1) \in F(\gterm{D})$. Тонкими линиями нарисованы ребра $\gterm{G}$, пунктиром~--- ребра $\gterm{G}^*$.}
        \label{picDoubleGraphFace}
    \end{figure}
    Голоморфные функции живут уже на вершинах $\gterm{D}$. Функция $f: V(\gterm{D}) \to \mathbb{C}$ называется дискретной голоморфной относительно весовой функции $\nu$, если для любой положительно ориентированной грани $(x_0, y_0, x_1, y_1) \in F(\gterm{D})$ (см. рисунок~\ref{picDoubleGraphFace}) выполняются дискретные уравнения Коши-Римана
    $$
        \frac{f(y_1) - f(y_0)}{f(x_1) - f(x_0)} = i\nu(x_0, x_1) = -\frac{1}{i\nu(y_0, y_1)}.
    $$

    Несложным вычислением проверяется, что ограничение дискретной голоморфной функции на любую из долей $V(\gterm{G})$, $V(\gterm{G}^*)$ является гармонической. Обратно, по любой гармонической функции на $V(\gterm{G})$ строится дискретная голоморфная на $V(\gterm{D})$, однозначная с точностью до прибавления константы на $V(\gterm{G}^*)$.

    Ребру из $\gterm{D}$ удобно присвоить в качестве метки тот координатный вектор $\mathbb{Z}^d$, в который переходит это ребро, а также ориентировать в сторону увеличения координаты.

    Отметим, что отображение $P$ любой квазикристаллической решетки в $\mathbb{Z}^d$, описанное в п. 3~\cite{discran}, имеет ровно такой же смысл, что и $\muld{n}$.

\subsection{Многоточечная волновая функция и дискретные уравнения Коши-Римана}
    Пусть дан планарный граф $\gterm{G}$. Построим по нему квад-граф $\gterm{D}$. Пусть существует отображение $\gterm{D}$ в целочисленную решетку $n: V(\gterm{D}) \to \mathbb{Z}^d$ для некоторого произвольного $d \ge 2$. Используя конечнозонный подход, мы построим по этому отображению весовую функцию $\nu$, через которую выписываются уравнения Коши-Римана на квад-графе $\gterm{D}$. Весовая функция, в свою очередь, определяется через обобщенные спектральные данные, о которых сейчас пойдет речь.

    Описанная ниже конструкция спектральных данных обобщает построения, сделанные в~\cite{giperbolDiscr} и в~\cite{4authors}. А именно, в указанных работах рассматривается случай квадратной решетки на плоскости и $d = 2$.

    Рассмотрим компактную, регулярную риманову поверхность $\Gamma$ рода $g$. Пусть на ней имеются следующие точки.
    \begin{itemize}
        \item Фиксированная точка $R_1$ на $\Gamma$ для нормировки волновой функции.
        \item Дивизор общего положения $\gamma_1, \dots, \gamma_g$.
        \item Коллекция из $d$ пар выделенных точек $A^+_1, A^-_1, A^+_2, A^-_2, \dots, A^+_d, A^-_d$. Все точки попарно различны.
    \end{itemize}
    По теореме Римана-Роха, для любого целочисленного вектора
    $$
        \muld{n} = (n_1, n_2, \dots, n_d) \in \mathbb{Z}^d
    $$
    существует единственная функция $\Psi(\muld{n}; \gamma)$, $\gamma \in \Gamma$, со следующими свойствами.
    \begin{enumerate}
        \item При каждом $\muld{n}$ функция $\Psi$ является мероморфной от $\gamma$.
        \item $\Psi$ имеет полюса не более первого порядка в точках $\gamma_1, \gamma_2, \dots, \gamma_g$.
        \item Для каждого $j = 1, 2, \dots, d$, $\Psi$ имеет полюс не более чем $n_j$ порядка в точке $A^+_j$ и нуль по крайней мере $n_j$ порядка в точке $A^-_j$.
        \item Выполняется условие нормировки $\Psi(\muld{n}; R_1) \equiv 1$.
    \end{enumerate}
    Функцию $\Psi$ называют волновой. Она естественно переводится на граф $\gterm{D}$:
    $$
        \Psi(p, \gamma) = \Psi(\muld{n}(p); \gamma), \quad p \in V(\gterm{D}).
    $$

\begin{figure}[h]
        \centering
        \includegraphics{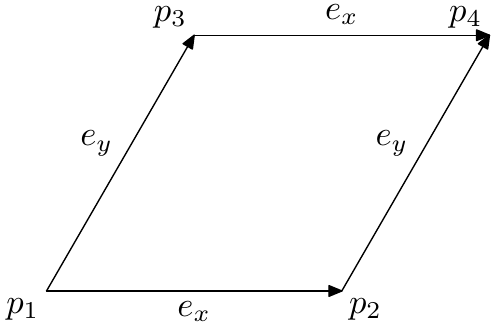}\\
        \caption{Грань $(p_1, p_2, p_4, p_3) \in F(\gterm{D})$.}
        \label{picGenericFace1}
    \end{figure}

    Рассмотрим произвольную (положительно ориентированную) грань $(p_1, p_2, p_4, p_3) \in F(\gterm{D})$. Пусть ребра этой грани были ориентированы в сторону вершин с большими номерами (см. рисунок~\ref{picGenericFace1}). Пусть ребро $(p_1, p_2)$ имеет метку $e_x$, а ребро $(p_1, p_3)$~--- метку $e_y$, $1 \le x, y \le d$, $x \ne y$.

    Заметим, что при данных общего положения у $\Psi(p_2, \gamma)$ полюс в $A^+_x$ на единицу большего порядка по сравнению с $\Psi(p_1, \gamma)$, а порядки остальных полюсов у них совпадают. Аналогично отличие видно между $\Psi(p_3, \gamma)$ и $\Psi(p_1, \gamma)$ по отношению к точке $A^\pm_y$. При переходе от $p_2$ или $p_3$ к $p_4$ наблюдается сходная картина. Сделанные замечания приводят нас к равенству
    \begin{gather}\label{eq4pointGeneric}
        \Psi(p_4, \gamma) + \alpha_1(p_1, p_2)\Psi(p_2, \gamma) + \alpha_2(p_1, p_3)\Psi(p_3, \gamma) + \alpha_3(p_1, p_4)\Psi(p_1, \gamma) = 0,
    \end{gather}
    где коэффициенты $\alpha_j$ не зависят от $\gamma$ и определяются по следующим формулам:
    \begin{gather}
        \alpha_1(p_1, p_2) = -\lim\limits_{\gamma \to A^+_x} \frac{\Psi(p_4, \gamma)}{\Psi(p_2, \gamma)}, \label{eqAlpha1} \\
        \alpha_2(p_1, p_3) = -\lim\limits_{\gamma \to A^+_y} \frac{\Psi(p_4, \gamma)}{\Psi(p_3, \gamma)}, \label{eqAlpha2} \\
        \alpha_3(p_1, p_4) = -1 - \alpha_1(p_1, p_2) - \alpha_2(p_1, p_3).
    \end{gather}
    Действительно, сумма \eqref{eq4pointGeneric} удовлетворяет всем условиям для $\Psi(p, \gamma)$ за исключением того, что в $R_1$ она обращается в нуль. Тогда по теореме Римана-Роха сумма равна нулю при любом $\gamma \in \Gamma$.

    Проделанные к этому моменту алгебро-геометрические построения аналогичны~\cite{giperbolDiscr}.

    Для того, чтобы уравнение~\eqref{eq4pointGeneric} привести к виду дискретного Коши-Римана, нам потребуются дополнительные условия на обобщенные данные рассеяния.
    \begin{Lemma}\label{lemmaSigmaOmega}
        Пусть на $\Gamma$ существует голоморфная инволюция $\sigma$ c двумя неподвижными точками $R_+ = R_1$ и $R_-$. Пусть спектральные данные обладают следующей симметрией.
        \begin{enumerate}
            \item Для $j = 1, \dots, d$ выполняется $\sigma A^+_j = A^-_j$.
            \item Существует мероморфный дифференциал $\Omega$ с двумя полюсами первого порядка в неподвижных точках $R_+$, $R_-$ и $2g$ нулями в $\gamma_1, \dots, \gamma_g$, $\sigma\gamma_1, \dots, \sigma\gamma_g$.
        \end{enumerate}
        Тогда
        \begin{gather}
            \Psi(p, R_-) = (-1)^{s(\muld{n}(p))}, \label{eqRminusValue} \\
            \alpha_1(x_0, y_0) = -\alpha_2(x_0, y_1), \quad \alpha_3(x_0, x_1) = -1,
        \end{gather}
        где $s(\muld{n}) = n_1 + n_2 + \dots + n_d$.
    \end{Lemma}
    \begin{Remark}
        Будем считать нормировку дифференциала $\Omega$ такой, что его вычеты в точках $R_+$, $R_-$ равны соответственно $\tfrac12$, $-\tfrac12$.
    \end{Remark}
    \begin{Remark}
        Условия леммы накладывают довольно сильные ограничения на спектральные данные.
    \end{Remark}

    Хочется отметить, что многоточечная волновая функция вместе с уравнением~\eqref{eq4pointGeneric} были построены еще в~\cite{discrPsiFunc} при более общих предположениях. В работе они используются для построения дискретного аналога решетки Дарбу--Егорова, размерность которой равна половине от количества неподвижных точек инволюции $\sigma$.

    Мы готовы определить весовую функцию $\nu: E(\gterm{G}) \sqcup E(\gterm{G}^*) \to \mathbb{C}$. Пусть условия леммы~\ref{lemmaSigmaOmega} выполнены. Вернемся к рассмотрению произвольной грани с описанной нумерацией вершин $(p_1, p_2, p_4, p_3) \in F(\gterm{D})$. Положим по определению
    \begin{gather}
        \nu(p_1, p_4) = \frac{1}{i \alpha_1(p_1, p_2)} = \frac{i}{\alpha_2(p_1, p_3)} \label{eqNu}, \\
        \nu(p_2, p_3) = \frac{1}{\nu(p_1, p_4)}.
    \end{gather}
    Перепишем равенство~\eqref{eq4pointGeneric} в терминах весовой функции
    $$
        \frac{\Psi(p_3, \gamma) - \Psi(p_2, \gamma)}{\Psi(p_4, \gamma) - \Psi(p_1, \gamma)} = i\nu(p_1, p_4) = -\frac{1}{i\nu(p_2, p_3)}.
    $$
    Рассмотрением всех 4 случаев взаимной ориентации ребер несложно показать, что данное равенство верно и без условия на ориентацию ребер. Можно сказать, что выполнением именно этого свойства продиктовано правило продолжения весовой функции с ребер $\gterm{G}$ на ребра двойственного графа $\nu(e^*) = 1/\nu(e)$.

    Таким образом, для произвольной грани ${(x_0, y_0, x_1, y_1) \in F(\gterm{D})}$ (рисунок~\ref{picDoubleGraphFace}) выполняется
    \begin{gather}\label{eqCauchyRiemannPsi}
        \frac{\Psi(y_1, \gamma) - \Psi(y_0, \gamma)}{\Psi(x_1, \gamma) - \Psi(x_0, \gamma)} = i\nu(x_0, x_1) = -\frac{1}{i\nu(y_0, y_1)}.
    \end{gather}
    По определению, волновая функция $\Psi(p, \gamma)$ является дискретной голоморфной на графе $\gterm{D}$ с весовой функцией $\nu$ при любом фиксированном $\gamma \in \Gamma$. Ограничение $\Psi(p, \gamma)$ на каждый из $\gterm{G}$, $\gterm{G}^*$ дает гармоническую функцию
    \begin{gather}\label{eqLaplacePsi}
        (L_x \Psi)(x_0, \gamma) = \sum\limits_{x \sim x_0} \nu(x_0, x)(\Psi(x, \gamma) - \Psi(x_0, \gamma)) = 0, \\
        (L_y \Psi)(y_0, \gamma) = \sum\limits_{y \sim y_0} \nu(y_0, y)(\Psi(y, \gamma) - \Psi(y_0, \gamma)) = 0.
    \end{gather}
    По умолчанию под $L$ мы будем подразумевать оператор $L_x$.

    В работе~\cite{discran} неоднократно отмечается особый интерес случая положительной весовой функции. Следующая лемма формулирует условия на обобщенные спектральные данные, достаточные для вещественности получаемой $\nu$.
    \begin{Lemma}\label{lemmaTau}
        Пусть выполнены условия леммы~\ref{lemmaSigmaOmega}. Пусть на $\Gamma$ существует антиголоморфная инволюция $\tau$ со следующими свойствами.
        \begin{enumerate}
            \item $\tau$ коммутирует с $\sigma$.
            \item $\tau R_+ = R_-$.
            \item Точки $A^+_1, A^-_1, \dots, A^+_d, A^-_d$ являются неподвижными для $\tau$.
            \item Дивизор $\gamma_1, \dots, \gamma_g$ переходит в себя под действием $\tau$ (но любая точка дивизора может переходить в отличную от себя).
        \end{enumerate}
        Тогда весовая функция $\nu$ принимает вещественные значения и выполняется
        \begin{gather}\label{eqRealityCondition}
            \Psi(p, \tau\gamma) = (-1)^{s(\muld{n}(p))}\overline{\Psi(p, \gamma)}.
        \end{gather}
    \end{Lemma}

    Проделанные к текущему моменту алгебро-геометрические построения аналогичны~\cite{4authors}. Инволюция $\tau$ также используется и в~\cite{discrPsiFunc} как условие вещественности решетки.

    \begin{Example}
        Покажем, что квазикристаллический случай полностью описывается спектральными данными, в которых $\Gamma$ является сферой Римана. Будем считать, что кроме отображения квад-графа $\muld{n}$ в $d$-мерную решетку задан набор попарно линейно независимых комплексных чисел $\{ \alpha_1, \dots, \alpha_d \}$. Каждому координатному вектору $\muld{e}_j \in \mathbb{Z}^d$ поставим в соответствие число $\alpha_j \in \mathbb{C}$. Этим действием мы задаем квазикристаллическое вложение квад-графа в комплексную плоскость, описанное в примере п.~3~\cite{discran}.

        Пусть $\Gamma = \overline{\mathbb{C}}$, инволюция $\sigma$ задается центральной симметрией $\sigma z = -z$ с неподвижными точками $R_- = 0$, $R_+ = \infty$, дифференциал $\Omega = -dz/2z$, гамма-дивизор пустой. В качестве выделенных точек $A^\pm_j$ на $\Gamma$ возьмем $\pm\alpha_j$. Волновая функция на сфере --- это дискретная экспонента
        $$
            \Psi(\muld{n}; z) = \prod\limits_{j = 1}^{d} \left(\frac{z + \alpha_j}{z - \alpha_j}\right)^{n_j}.
        $$
        Несложно проверить, что построенная по $\Psi$ весовая функция равна
        $$
            \nu(p_2, p_3) = \frac{1}{\nu(p_1, p_4)} = i\frac{\alpha_y - \alpha_x}{\alpha_y + \alpha_x}.
        $$
        Она совпадает с построенной по квазикристаллическому вложению: значение $i\nu$ на любом параллелограмме равно отношению его диагоналей. В~\cite{discran} также доказано, что это равносильно интегрируемости уравнения Коши-Римана в смысле <<3D-совместности>>.

        В случае вещественных весов из последней формулы следует, что все $\alpha_j$ равны между собой по абсолютной величине: $\abs{\alpha_j}^2 = C^2$. Определим антиголоморфную инволюцию $\tau z = C / \overline{z}$, тогда все $\alpha_j$ будут для нее неподвижными точками и условия леммы~\ref{lemmaTau} будут выполнены. Таким образом, в квазикристаллическом случае достаточное условие вещественности является еще и необходимым.
    \end{Example}

\subsection{Рост волновой функции}
\label{sectionPsiGrowth}
    Вопрос о том, как ведет себя $\abs{\Psi(\muld{n}, \gamma)}$ при фиксированном $\gamma$, очень важен для оценки роста функции Грина. Для формулировки и доказательства теоремы нам потребуются некоторые понятия теории римановых поверхностей.
    \def\K1{\sum\limits_{k=1}^{g}\Abel{\gamma_k} - \vec{K}}
    \newcommand\FTheta[1]{\theta\left(\Abel{#1} + \sum\limits_{j = 1}^{d} n_j\vec{\Delta}_j - \K1\right)}
    \newcommand\Expint[1]{\int\limits_{R_+}^{\gamma}\Omega(#1^+, #1^-)}
    \newcommand\ExpintN{\exp\left( \sum\limits_{j = 1}^{d} n_j\Expint{A_j} \right)}
    \newcommand\vdelta{\vec{\Delta}}

    Выберем на $\Gamma$ канонический базис циклов $a_1, \dots, a_g, b_1, \dots, b_g$ и базис голоморфных дифференциалов $\omega_1, \dots, \omega_g$, нормированный следующим образом:
    $$
        \oint\limits_{a_k} \omega_j = \delta_{jk}.
    $$
    Нам понадобится тета-функция Римана поверхности $\Gamma$, которая определяется рядом
    $$
        \theta(z|B) = \sum\limits_{N \in \mathbb{Z}^g} \exp\left( \pi i \ScalarP{BN}{N} + 2\pi i \ScalarP{N}{z} \right), \quad z \in \mathbb{C}^g,
    $$
    где $\ScalarP{\cdot}{\cdot}$ --- евклидово скалярное произведение, а $B$ --- матрица $b$-периодов голоморфных дифференциалов
    $$
        \oint\limits_{b_k} \omega_j = B_{jk},
    $$
    мнимая часть которой оказывается положительно определена. Зададим отображение Абеля как
    \begin{gather}\label{defAbel}
        \Abel{\gamma} = \left( \int\limits_{R_+}^\gamma \omega_1, \dots, \int\limits_{R_+}^\gamma \omega_g \right).
    \end{gather}
    Напомним, что это корректно определенное отображение $\Gamma \xrightarrow{A} J(\Gamma)$, где $J(\Gamma)$ --- многообразие Якоби, ${J(\Gamma) = \mathbb{C}^g / \{M + BN\}}$ для $M, N \in \mathbb{Z}^g$.

    Для двух различных точек $P$, $Q$ римановой поверхности существует мероморфный дифференциал $\Omega(P, Q)$ с полюсами первого порядка в $P$, $Q$ и вычетами $-1$ и $1$ соответственно, не имеющий других особенностей. Мы добавим условие равенства нулю по всем $a$-циклам, благодаря которому $\Omega(P, Q)$ определяется однозначно. Он противоположен соответствующему нормированному абелеву дифференциалу третьего рода.

    По аналогии с 5.2~\cite{4authors}, для волновой функции $\Psi$ можно написать явную формулу, верную при любых $\muld{n} \in \mathbb{Z}^d$:
    \begin{gather}\label{psiExplicitFormula}
        \Psi(\muld{n}, \gamma) = \ExpintN \times\\
        \times \frac{\FTheta{\gamma}}{\theta\left( \Abel{\gamma} - \K1 \right)} \times \frac{\theta\left( \Abel{R_+} - \K1 \right)}{\FTheta{R_+}},\notag
    \end{gather}
    где
    $$
        \vdelta_j = \Abel{A_j^-} - \Abel{A_j^+}
    $$
    и пути во всех интегралах берутся одинаковыми. Проверим, что~\eqref{psiExplicitFormula} задаёт однозначную на $\Gamma$ функцию. Если путь до фиксированного $\gamma$ изменяется на некоторый цикл, гомологичный
    $$
        \sum_{j = 1}^{g} (N_j a_j + M_j b_j), \quad \vec{N}, \vec{M} \in \mathbb{Z}^g,
    $$
    то отношение $\theta$-функций умножится на
    $$
        t = \exp\left(-2\pi i \ScalarP{\vec{M}}{\sum\limits_{j = 1}^{g} n_j\vdelta_j}\right).
    $$
    Из теории римановых поверхностей нам известно, что
    \begin{gather}\label{omegaRelation}
        \oint\limits_{b_k} \Omega(A_j^+, A_j^-) = 2\pi i \int\limits_{A_j^+}^{A_j^-} \omega_k,
    \end{gather}
    а следовательно, экспонента умножится на $t^{-1}$.

    Пусть $\Gamma$ является М-кривой, то есть инволюция $\tau$ имеет $g + 1$ неподвижный овал $a_1, a_2, \dots, a_g, c$.
    \begin{Theorem}\label{thPsiGrowth}
        Пусть $\Gamma$ является М-кривой, выделенные точки $A_j^\pm$, $j = 1, \dots, d$, попадают на овал $c$, на остальные овалы попадает по одной точке $\gamma$-дивизора: $\gamma_k \in a_k$, $k = 1, \dots, g$. Тогда канонический базис циклов и пути интегрирования на $\Gamma$ можно выбрать таким образом, что для любого фиксированного $\gamma \in \Gamma \setminus (a_1 \cup \dots \cup a_g \cup c)$ выполняется неравенство при всех $\muld{n} \in \mathbb{Z}^d$:
        \begin{gather}
            \abs{\Psi(\muld{n}, \gamma)} \le R(\gamma)\abs{\ExpintN}, \label{growthRaw}
        \end{gather}
        где $R: \Gamma \to \mathbb{R}$ --- гладкая на $\Gamma \setminus (a_1 \cup \dots \cup a_g \cup c)$ функция.
    \end{Theorem}
    Другими словами, почти всех $\gamma \in \Gamma$ рост абсолютной величины $\Psi(\muld{n})$ зависит только от $\Omega(A_j^+, A_j^-)$, $j = 1, \dots, d$.
    \begin{Proof}
        Благодаря расположению $\gamma_k$ все нули $\Psi(\muld{n}, \gamma)$ при любых $\muld{n} \in \mathbb{Z}^d$ располагаются только на неподвижных овалах $a_1, \dots, a_g, c$. Действительно, на каждом из $a_k$ ($k = 1, \dots, g$), функция $\Psi(\muld{n}, \gamma)$ вещественная или чисто мнимая~\eqref{eqRealityCondition} и имеет полюс первого порядка. Тогда на $a_k$ найдется и нуль по крайней мере первого порядка. Степень дивизора $\Psi$
        $$
            \sum\limits_{j = 1}^{d} n_j(A_j^- - A_j^+) - \sum\limits_{k = 1}^{g} \gamma_k
        $$
        равна $(-g)$ и по построению у $\Psi(\muld{n}, \gamma)$ нет полюсов вне точек этого дивизора. Следовательно, все нули на $a_k$ имеют первый порядок и больше на $\Gamma$ нулей у $\Psi(\muld{n}, \gamma)$ нет.

        Рассмотрим явную формулу~\eqref{psiExplicitFormula}. Пусть $\gamma \in \Gamma \setminus (a_1 \cup \dots \cup a_g \cup c)$, тогда ни одна из $\theta$-функций не обращается в нуль. Мы докажем существование гладких $R_{min}(\gamma) > 0$ и $R_{max}(\gamma) > 0$, таких что для любых $\muld{n} \in \mathbb{Z}^d$ выполняется
        $$
            R_{min}(\gamma) \le \abs{\FTheta{\gamma}} \le R_{max}(\gamma).
        $$
        Искомая оценка будет выполняться при
        $$
            R(\gamma) = \frac{R_{max}(\gamma)}{R_{min}(R_+)} \frac{\abs{\theta\left(\Abel{R_+} - \K1\right)}}{\abs{\theta\left(\Abel{\gamma} - \K1\right)}}.
        $$

        Возьмем в качестве $a$-циклов канонического базиса неподвижные овалы $\tau$ с точками $\gamma$-дивизора $a_1, \dots, a_g$. Благодаря такому выбору мы получаем целый ряд свойств.

        Для каждого $k = 1, \dots, g$ дифференциал $\overline{\tau \omega_k}$ является голоморфным и имеет ту же нормировку, что и $\omega_k$. Следовательно, $\tau\omega_k = \overline{\omega}_k$ и $\omega_k$ принимает вещественные значения на неподвижных овалах $\tau$.

        Вещественной частью многообразия Якоби $\Re J(\Gamma)$ назовем подмножество $J(\Gamma)$ классов эквивалентности с вещественными представителями $\vec{x} + B\vec{M}$, где $\vec{x} \in \mathbb{R}^g$, $\vec{M} \in \mathbb{Z}^g$. По построению, вещественная часть является замкнутым множеством.

        Вспомним, что при изменении $\muld{d}$ аргументы $\theta$-функций изменяются на $\vdelta_j = \Abel{A_j^-} - \Abel{A_j^+}$, $j = 1, \dots, d$. Тогда из вещественности $\omega_k$ на неподвижных овалах и определения
        \begin{gather*}
            (\vdelta_j)_k = \int\limits_{A_j^+}^{A_j^-} \omega_k
        \end{gather*}
        следует $\vdelta_j \in \Re J(\Gamma)$, так как от вещественного вектора они могут отличаться только на периоды многообразия Якоби.

        Фиксируем $\lambda \in \Gamma \setminus (a_1 \cup \dots \cup a_g \cup c)$ и рассмотрим множество всех значений аргументов рассматриваемой $\theta$-функции при различных $\muld{n} \in \mathbb{Z}^d$:
        $$
            V(\lambda) = \left\{\left. \Abel{\lambda} + \sum\limits_{j = 1}^{d} n_j\vdelta_j - \K1 \right| \muld{n} \in \mathbb{Z}^d \right\}.
        $$
        Докажем, что замыкание $V(\lambda)$ в $J(\Gamma)$ не содержит нулей $\theta$-функции. Пусть такой нуль $z \in J(\Gamma)$ все-таки нашелся. Тогда разность
        $$
            z - \left(\Abel{\lambda} - \K1\right)
        $$
        сколь угодно приближается суммой
        $$
            \sum\limits_{j = 1}^{d} n_j\vdelta_j \in \Re J(\Gamma)
        $$
        и по замкнутости сама принадлежит $\Re J(\Gamma)$. Следовательно, найдется такая $\lambda_0 \in \Gamma$, $\tau\lambda_0 = \lambda_0$, что на $J(\Gamma)$ выполняется равенство
        $$
            z = \Abel{\lambda_0} - \K1,
        $$
        откуда следует $\Abel{\lambda_0} - \Abel{\lambda} \in \Re J(\Gamma)$. Воспользуемся теперь возможностью выбрать пути интегрирования и добьемся вещественности последней разности: $\Abel{\lambda_0} - \Abel{\lambda} \in \mathbb{R}^g$. Из $\tau\omega_j = \overline{\omega}_j$ вытекает
        $$
            \Abel{\lambda_0} - \Abel{\tau\lambda} = \overline{\Abel{\lambda_0} - \Abel{\lambda}},
        $$
        а из вещественности правой части $\Abel{\lambda} = \Abel{\tau\lambda}$. Поскольку $\tau\lambda \ne \lambda$, такое может быть только на сфере $g = 0$, где доказываемая оценка тривиальна.

        Из отсутствия нулей в замыкании $V(\lambda) \subset J(\Gamma)$ и компактности последнего следует существование искомых $R_{min}(\lambda)$, $R_{max}(\lambda)$ для всех $\lambda \notin \left( a_1 \cup \dots \cup a_h \cup c \right)$, этим и завершается доказательство.
    \end{Proof}
    \begin{Remark}\label{remarkPsiGrowth}
        Выбор путей интегрирования в точности соответствует случаю ${\vec{\Delta}_j \in \mathbb{R}^g}$, $j = 1, \dots, d$, поэтому по~\eqref{omegaRelation} интегралы от $\Omega(A_j^+, A_j^-)$ по любому циклу являются вещественными.
    \end{Remark}
    \begin{Remark}
        По всей видимости, оценка~\eqref{growthRaw} выполняется почти всюду и в более общем случае, когда $\Gamma$ не является M-кривой. Но строгое доказательство требует более серьезной техники. Эта задача --- тема для дальнейших исследований.
    \end{Remark}

\subsection{Квазиимпульсы}
    Дифференциалы квазиимпульсов $dp_j$, $j = 1, \dots, d$, определяются по аналогии с~\cite{KdVandFiniteZone}. А именно, это мероморфные дифференциалы третьего рода; $dp_j$ имеет вычеты $i$, $-i$ в точках $A_j^+$, $A_j^-$ соответственно. Дифференциалы квазиимпульсов однозначно определяются условием вещественности интегралов по всем контурам. Сами квазиимпульсы определяются как
    \begin{gather}
        p_j(\gamma) = \int\limits_{R_+}^{\gamma} dp_j
    \end{gather}
    и являются многозначными на $\Gamma$, однако их мнимые части $\Im p_j(\gamma)$ уже являются однозначными на $\Gamma$.

    Из замечания~\ref{remarkPsiGrowth} и единственности дифференциалов квазиимпульсов следует, что при выборе канонического базиса циклов и путей интегрирования как в теореме~\ref{thPsiGrowth} выполняется $\Omega(P^+, P^-) = -idp_m$, $\Omega(Q^+, Q^-) = -idp_n$. Поэтому оценка~\eqref{growthRaw} может быть переписана в терминах квазиимпульсов:
    \begin{gather}\label{eqGrowthPsi}
        \abs{\Psi(\muld{n}, \gamma)} \le R(\gamma) e^{\ScalarP{\muld{n}}{\Im\muld{p}(\gamma)}},
    \end{gather}
    где
    $$
        \muld{p}(\gamma) = (p_1(\gamma), \dots, p_d(\gamma)), \quad \ScalarP{\muld{n}}{\Im\muld{p}(\gamma)} = \sum\limits_{j = 1}^{d} n_j \Im p_j(\gamma).
    $$
    Отметим, что поскольку и левая часть, и квазиимпульсы уже не зависят от выбора базиса или путей интегралов, то функция $R(\gamma)$ также не зависит от них.

    Оценка абсолютной величины двойственной волновой функции получается заменой $\gamma$ на $\sigma\gamma$
    $$
        \abs{\Psi^+(\muld{n}, \gamma)} \le R(\sigma\gamma) e^{\ScalarP{\muld{n}}{\Im\muld{p}(\sigma\gamma)}}.
    $$
    Дифференциал $-dp_j(\sigma\gamma)$ имеет полюса в $A_j^+$, $A_j^-$ с вычетами соответственно $+i$, $-i$, а также интеграл от него по любому контуру является вещественным. Следовательно, $dp_j(\sigma\gamma) = -dp_j$. Поэтому последнее неравенство можно переписать в виде
    \begin{gather}\label{eqGrowthPsiPlus}
        \abs{\Psi^+(\muld{n}, \gamma)} \le R(\sigma\gamma) e^{-\ScalarP{\muld{n}}{\Im\muld{p}(\gamma)}}.
    \end{gather}

    Рассмотрим вещественную линейную комбинацию дифференциалов квазиимпульсов:
    $$
        dp_\muld{k} = k_1 dp_1 + \dots + k_d dp_d, \quad \muld{k} \in \mathbb{R}^d \setminus \{ \muld{0} \}.
    $$
    Построим по ней множество точек на $\Gamma$:
    $$
        C_\muld{k}(\lambda) =  \{ \gamma: \Im p_{\muld{k}}(\gamma) = \Im p_\muld{k}(\lambda) \}.
    $$

    Эти контуры мы будем использовать при построении функции Грина для контроля роста волновой функции. Такого рода контуры возникли ещё в работе Кричевера и Новикова~\cite{KN}.

    \begin{Example} Продолжим рассмотрение квазикристаллического случая при $g = 0$. В качестве дифференциалов квазиимпульсов подходят
        $$
            dp_j = \frac{i dz}{z - \alpha_j} - \frac{i dz}{z + \alpha_j}, \quad j = 1, \dots d.
        $$
        Действительно, мнимые части квазиимпульсов получаются однозначными:
        $$
            p_j = i\ln\left(\frac{z - \alpha_j}{z + \alpha_j}\right), \quad \Im p_j = \ln\left|\frac{z - \alpha_j}{z + \alpha_j}\right|.
        $$
        На сфере Римана контуры $\Im p_\muld{k} = \const$ представляют собой эллипсы, превращающиеся в окружности при равенстве нулю одной из компонент.

        Оценки~\eqref{eqGrowthPsi} и~\eqref{eqGrowthPsiPlus} в случае сферы обращаются в равенства с $R \equiv 1$.
    \end{Example}

    \begin{Lemma}\label{lemmaMainCountourRegularity}
        Почти для всех $\lambda \in \Gamma \setminus \{ A_j^+, A_j^- \}$ множество $C_\muld{k}(\lambda)$ при любом $\muld{k} \in \mathbb{R}^d \setminus \{\muld{0}\}$ является объединением конечного числа непрерывных замкнутых кривых. Ориентация на нем корректно задается условием $\Re dp_\muld{k}(\gamma) > 0$, получаемый при этом контур оказывается гомологичным точке. Кроме того,
        \begin{gather}\label{eqOmegaCkZero}
            \oint\limits_{C_\muld{k}(\lambda)} \Omega = 0,
        \end{gather}
        то есть, точки $R_+$, $R_-$ одинаково обмотаны этим контуром.
    \end{Lemma}
    \begin{Proof}
        Функция $\Im p_\muld{k}$ является гармонической и множество ее нулей не имеет на $\Gamma$ внутренних точек. Благодаря этому почти для всех $\lambda$ рассматриваемое множество точек, находящихся на одном уровне с $\lambda$, является непрерывным. Конечность и замкнутость кривых следует из компактности $\Gamma$. Контур $C_\muld{k}(\lambda)$ с указанной ориентацией гомологичен нулю как граница подмногообразия $\{ \gamma: \Im p_\muld{k}(\gamma) \le \Im p_\muld{k}(\lambda) \}$.

        Проверим второе утверждение. Зададим любую ориентацию на контуре. Из соображений единственности следует $\tau\Omega = -\overline{\Omega}$. Кроме того, несложно проверить $\Im p_k(\gamma) = \Im p_k(\tau\gamma)$, откуда следует $\tau(C_\muld{k}(\lambda)) = -C_\muld{k}(\lambda)$. Получаем
        $$
            \oint\limits_{C_\muld{k}(\lambda)} \Omega = \oint\limits_{-C_\muld{k}(\lambda)} -\overline{\Omega},
        $$
        то есть вещественность значения интеграла~\eqref{eqOmegaCkZero}. Но мы знаем, что у $\Omega$ всего два полюса в $R_+$, $R_-$ с вычетами $\pm\tfrac12$, поэтому значение интеграла~\eqref{eqOmegaCkZero} должно быть мнимым. Из этого следует его равенство нулю и равенство коэффициентов, с которыми вычеты в $R_+$, $R_-$ входят в значение интеграла.
    \end{Proof}

    \begin{Lemma}\label{lemmaCountourHomology}
        Рассмотрим два произвольных вектора $\muld{k}, \muld{k'} \in \mathbb{R}^d \setminus \{0\}$. Пусть $J \subset \{ 1, \dots, d \}$~--- произвольное подмножество индексов $j$, таких что $k_j k'_j > 0$. Тогда почти для всех $\lambda \in \Gamma$ контуры $C_\muld{k}(\lambda)$, $C_\muld{k'}(\lambda)$ гомологичны в $\Gamma \setminus \{ A^\pm_j: j \in J\}$.
    \end{Lemma}
    \begin{Proof}
        Достаточно разобрать случай $\muld{k'} = \muld{k} + \muld{e_i}$ для произвольного $i = 1, \dots, d$. Для этого рассмотрим деформацию одного контура в другой $C_t(\lambda) = C_\muld{k + te_i}(\lambda)$, $t \in [0, 1]$. Ни один из контуров деформации не проходит через выколотые точки $\{ A^\pm_j: j \in J \}$, поскольку $\Im p_\muld{k + te_i}$ обращается в $\pm\infty$ в этих точках (и $\Im p_\muld{k + te_i}(\lambda)$ конечно). Следовательно, деформация $C_t(\lambda)$ непрерывна в $\Gamma$ без указанных точек, исходные контуры гомотопны, а значит и гомологичны.
    \end{Proof}

\subsection{Функция Грина оператора Лапласа}
    Рассмотрим оператор Лапласа $L = L_x$, действующий на $\gterm{G}$ по формуле~\eqref{eqLaplacePsi}. Нас интересует такая функция $G(x, \tilde x, \lambda)$, $x \in V(\gterm{G})$, $\tilde x \in V(\gterm{G})$, что для (почти) любого фиксированного $\lambda \in \Gamma$ выполняется
    \begin{gather}\label{eqGreenEquality}
        LG =
        \begin{cases}
            1, & \text{если $x = \tilde x$},\\
            0 & \text{иначе},
        \end{cases}
    \end{gather}
    где
    $$
        (LG)(x, \tilde x, \lambda) = \sum\limits_{x_1 \sim x} \nu(x, x_1) (G(x_1, \tilde x, \lambda) - G(x, \tilde x, \lambda)).
    $$
    Забегая вперед, скажем, что почти при всех $\lambda$ для найденной функции $G$ выполнено условие роста
    \begin{gather}\label{eqGrowthGreen}
        \abs{G(x, \tilde x, \lambda)} \le R_1(\lambda) e^{\ScalarP{\muld{n}(x) - \muld{n}(\tilde x)}{\Im p_{\muld{n}(x) - \muld{n}(\tilde x)}(\lambda)}},
    \end{gather}
    где $R_1: \Gamma \to \mathbb{R}$ --- гладкая в точках выполнения неравенства. Другими словами, почти всюду рост абсолютной величины $G$ такой же, как и $\Psi$.

    Вслед за авторами упомянутых работ, постараемся найти выражение функции Грина в виде интеграла от $\Psi(x, \gamma)\Psi^+(\tilde x, \gamma)\Omega(\gamma)$ по контуру $C_{\muld{n}(x) - \muld{n}(\tilde x)}(\lambda)$. Для этого исследуем функцию
    $$
        H(x, \tilde x, \lambda) = \oint\limits_{C_{\muld{n}(x) - \muld{n}(\tilde x)}(\lambda)} \Psi(x, \gamma)\Psi^+(\tilde x, \gamma)\Omega(\gamma), \quad x \ne \tilde x,
    $$
    положив $H(x, x, \lambda) = 0$. Такое доопределение выбрано для согласования с равенством нулю дифференциала $\Psi(x, \gamma)\Psi^+(x, \gamma)\Omega(\gamma)$ по любому C-контуру рассматриваемого вида. По лемме~\ref{lemmaMainCountourRegularity}, почти при всех $\lambda$ приведенная формула корректно задает $H$ для всех вершин графа $\gterm{G}$.

    Напомним, что дифференциал $\Psi(x, \gamma)\Psi^+(\tilde x, \gamma)\Omega(\gamma)$ имеет следующий набор особенностей:
    \begin{itemize}
        \item Полюс порядка $n_j(x) - n_j(\tilde x)$ в $A^+_j$ и нуль такого же порядка в $A^-_j$, $j = 1, \dots, d$.
        \item Полюса первых порядков в $R_\pm$ с вычетами $\pm\tfrac12 (-1)^{s(\muld{n}(x)) - s(\muld{n}(\tilde x))}$.
    \end{itemize}

    \begin{Theorem}\label{thHeqZero}
        Почти для всех $\lambda \in \Gamma$ и любых $x \in V(\gterm{G})$, $\tilde x \in V(\gterm{G})$ выполняется $(LH)(x, \tilde x, \lambda) \equiv 0$.
    \end{Theorem}
    Для начала доказательства распишем
    $$
        (LH)(x, \tilde x, \lambda) = \sum\limits_{x_1 \sim x} \nu(x, x_1) \left( \oint\limits_{C_{\muld{n}(x_1) - \muld{n}(\tilde x)}(\lambda)} \Psi(x_1, \gamma) \Psi^+(\tilde x, \gamma) \Omega(\gamma) - \oint\limits_{C_{\muld{n}(x) - \muld{n}(\tilde x)}(\lambda)} \Psi(x, \gamma) \Psi^+(\tilde x, \gamma) \Omega(\gamma) \right)
    $$

    \begin{Lemma}\label{lemmaIntegrationWithinEdge}
        Для любой грани $(p_1, p_2, p_3, p_4)$ графа $\gterm{D}$ найдется вершина $p_k$, такая что для любого $j = 1, \dots, 4$ выполняется
        $$
            \oint\limits_{C_{\muld{n}(p_j) - \muld{n}(\tilde x)}(\lambda)} \Psi(p_j, \gamma)\Psi^+(\tilde x) \Omega(\gamma) =
            \oint\limits_{C_{\muld{n}(p_k) - \muld{n}(\tilde x)}(\lambda)} \Psi(p_j, \gamma)\Psi^+(\tilde x) \Omega(\gamma).
        $$
    \end{Lemma}
    Действительно, по лемме~\ref{lemmaMainCountourRegularity} достаточно учесть только особенности в $A^\pm_j$, $j = 1, \dots, d$, поскольку вычеты в $R_\pm$, равные $\pm\tfrac12$, вносят в сумму нулевой вклад. Искомая $p_k$ --- это вершина с лексикографически максимальным $(|n_1(p_k) - n_1(\tilde x)|, \dots, |n_d(p_k) - n_d(\tilde x)|)$.

    Обозначим соответствующий контур $C_{\muld{n}(p_k) - \muld{n}(\tilde x)}(\lambda)$ через $C_{p_1, p_4, \tilde x}(\lambda)$. Тогда по лемме
    $$
        (LH)(x, \tilde x, \lambda) = \sum\limits_{x_1 \sim x} \oint\limits_{C_{x_1, x, \tilde x}(\lambda)}
        \nu(x, x_1) (\Psi(x_1, \gamma)  - \Psi(x, \gamma)) \Psi^+(\tilde x, \gamma) \Omega(\gamma).
    $$
    Воспользуемся дискретной голоморфностью $\Psi$. Кроме того, поменяем порядок суммирования. Для этого рассмотрим две соседние грани в графе $\gterm{D}$: $(x, y_0, x_0, y_1)$, $(x, y_1, x_1, y_2)$, имеющие общие вершины $x$ и $y_1$. Под суммированием по всем $y_1 \sim x$ будем подразумевать суммирование по всем таким граням. Тогда
    \begin{gather}\label{eqPreHeqZero}
        (LH)(x, \tilde x, \lambda) = \sum\limits_{y_1 \sim x} i \left( \oint\limits_{C_{x, x_2, \tilde x}(\lambda)} - \oint\limits_{C_{x, x_1, \tilde x}(\lambda)} \right) \Psi(y_1, \gamma)\Psi^+(\tilde x, \gamma) \Omega(\gamma).
    \end{gather}

    \begin{Lemma}
        Вычеты полюсов из $A^\pm_j$, $j = 1, \dots, d$ дают одинаковые вклады в интегралы дифференциала $\Psi(y_1, \gamma)\Psi^+(\tilde x, \gamma) \Omega(\gamma)$ по $C_{x, x_2, \tilde x}(\lambda)$ и по $C_{x, x_1, \tilde x}(\lambda)$. Другими словами, при вычислении их разности достаточно учитывать только полюса $R_+$, $R_-$.
    \end{Lemma}
    \begin{Proof}
        Рассмотрим подмножество $J \subset \{1, \dots, d\}$ индексов, таких что $n_j(y_1) \ne n_j(\tilde x)$, то есть набор индексов тех точек $A^\pm_j$, в которых дифференциал $\Psi(y_1, \gamma)\Psi^+(\tilde x, \gamma)\Omega(\gamma)$ имеет полюс. Несложно убедиться, что по построению $J$ удовлетворяет условиям леммы~\ref{lemmaCountourHomology} по отношению к контурам $C_{x, x_2, \tilde x}(\lambda)$ и по $C_{x, x_1, \tilde x}(\lambda)$, поскольку они построены на вершинах из соседних граней. Отсюда вытекает гомологичность на $\Gamma \subset \{A^\pm_j: j \in J\}$ и утверждение леммы.
    \end{Proof}
    Осталось учесть вычеты в $R_+$, $R_-$. Заметим, что для любого $y_1 \sim x$ вычеты дифференциала $\Psi(y_1, \gamma)\Psi^+(\tilde x, \gamma)\Omega(\gamma)$ в $R_+$, $R_-$ равны $\tfrac12$. Поскольку каждый из контуров $C_{p_1,p_4,\tilde x}(\lambda)$ входит в сумму~\eqref{eqPreHeqZero} дважды с разными знаками, то соответствующие слагаемые взаимно уничтожатся.

    Таким образом, $LH \equiv 0$ и теорема~\ref{thHeqZero} доказана.

    \begin{Theorem}
        В условиях теоремы~\ref{thPsiGrowth} cледующими формулами почти при любых $\lambda \in \Gamma$ определяется Функция Грина оператора $L$ на графе $\gterm{G}$, имеющая асимптотику~\eqref{eqGrowthGreen}
        \begin{gather}\label{eqGreenFormula}
            G(x, \tilde x, \lambda) = \frac{1}{k(\tilde x)}\oint\limits_{C_{\muld{n}(x) - \muld{n}(\tilde x)}(\lambda)} \Psi(x, \gamma)\Psi^+(\tilde x, \gamma)\Omega(\gamma), \quad x \ne \tilde x,
        \end{gather}
        $$
            G(\tilde x, \tilde x, \lambda) = 1,
        $$
        где
        $$
            k(\tilde x) = \sum\limits_{\tilde x_1 \sim \tilde x} \nu(\tilde x, \tilde x_1).
        $$
    \end{Theorem}
    \begin{Proof}
        Из предыдущей теоремы следует, что $LG(x, \tilde x, \lambda) = 0$, если $x \ne \tilde x$. Из нее же
        $$
            (LG)(\tilde x, \tilde x, \lambda) = \frac{1}{k(\tilde x)} \sum\limits_{\tilde x_1 \sim \tilde x} \nu(\tilde x, \tilde x_1) (G(\tilde x_1, \tilde x, \lambda) - G(\tilde x, \tilde x, \lambda)) = -\frac{G(\tilde x, \tilde x, \lambda)}{k(\tilde x)}\sum\limits_{\tilde x_1 \sim \tilde x} \nu(\tilde x, \tilde x_1) = 1.
        $$

        Проверим теперь, что построенная $G$ удовлетворяет условиям роста~\eqref{eqGrowthGreen}. Значение $G$ не изменится, если из контура интегрирования мы выкинем все неподвижные точки инволюции $\tau$ (множество меры нуль). Обозначим этот контур через $D(x, \tilde x, \lambda)$. Оценим интеграл стандартным образом
        $$
            \abs{G(x, \tilde x, \lambda)} \le \frac{1}{k(\tilde x)} \sup\limits_{\gamma \in D(x, \tilde x, \lambda)} \abs{\Psi(x, \gamma)\Psi^+(\tilde x, \gamma)} \oint\limits_{D(x, \tilde x, \lambda)} \abs{\Omega(\gamma)}
        $$
        По теореме~\ref{thPsiGrowth}, для всех точек $\gamma \in D(x, \tilde x, \lambda)$ выполняются неравенства~\eqref{eqGrowthPsi},~\eqref{eqGrowthPsiPlus}. По определению контура $C_{\muld{n}(x) - \muld{n}(\tilde x)}(\lambda)$, в этих неравенствах можно заменить $\gamma$ на $\lambda$.

        Далее, функция
        $$
            R_\Omega(\lambda) = \sup\limits_{x, \tilde x} \oint\limits_{D(x, \tilde x, \lambda)} \abs{\Omega(\gamma)}
        $$
        принимает конечные значения. Действительно, все контуры интегрирования параметризуются вектором $\muld{n}(x) - \muld{n}(\tilde x)$ с точностью до ненулевого множителя, а потому лежат в некотором компакте в $\mathbb{R}P^d$. В любой точке этого компакта значение интеграла конечно, тогда и $\sup$ по нему конечен.

        Из всего сказанного следует
        $$
            \abs{G(x, \tilde x, \lambda)} \le R_1(\lambda) e^{\ScalarP{\muld{n}(x) - \muld{x}(\tilde x)}{\Im p_{\muld{n}(x) - \muld{n}(\tilde x)}(\gamma)}},
        $$
        где
        $$
            R_1(\lambda) = \frac{1}{k(\tilde x)} R_\Omega(\lambda) \sup\limits_{\gamma \in D(x, \tilde x, \lambda)} (R(\gamma)R(\sigma\gamma))
        $$
        $R_1(\lambda)$~--- гладкая функция $\Gamma \setminus \{ a_1, \dots, a_g, c \} \to \mathbb{R}$, что и требовалось.
    \end{Proof}

\end{document}